\documentclass[8pt,twocolumn]{article}

\usepackage[numbers]{natbib}
\usepackage{graphicx}
\usepackage{caption}
\usepackage{subcaption}
\usepackage{abstract}

\usepackage{tikz}
\usetikzlibrary{arrows,shapes,snakes,automata,backgrounds,petri,through,positioning}
\usetikzlibrary{intersections}
\usetikzlibrary{knots}
\usepackage{pgfplots}
\pgfplotsset{compat=1.18}

\newtheorem{proposition}{Proposition} 
\newtheorem{conjecture}{Conjecture}

\newtheorem{example}{Example}

\title{Stable Andrews-Curtis trivialization of AK(3) revisited. \\ A case study using automated deduction.  }  
\author{Alexei Lisitsa\\ University of Liverpool\\ email: a.lisitsa@liverpool.ac.uk } 
\date{}

\begin{document}
\maketitle
    \begin{abstract}
    Recent work by Shehper et al. (2024) demonstrated that the well-known Akbulut-Kirby 
AK(3) balanced presentation of the trivial group is stably AC-equivalent to the trivial presentation. This result eliminates 
AK(3) as a potential counterexample to the stable Andrews-Curtis conjecture. In this paper, we present an alternative proof of this result using an automated deduction approach. We provide several transformation sequences, derived from two different proofs generated by the automated theorem prover Prover9, that certify the stable AC-equivalence of 
AK(3) and the trivial presentation. We conclude by proposing  a challenge to develop computational methods for searching stable AC-transformations.
    \end{abstract}

\maketitle

\newcommand{\abstractText}{\noindent
Recent work by Shehper et al. (2024) demonstrated that the well-known Akbulut-Kirby 
AK(3) balanced presentation of the trivial group is stably AC-equivalent to the trivial presentation. This result eliminates 
AK(3) as a potential counterexample to the stable Andrews-Curtis conjecture. In this paper, we present an alternative proof of this result using an automated deduction approach. We provide several transformation sequences, derived from two different proofs generated by the automated theorem prover Prover9, that certify the stable AC-equivalence of 
AK(3) and the trivial presentation. We conclude by proposing  a challenge to develop computational methods for searching stable AC-transformations.
}



%
%

\section{Introduction}
The Andrews-Curtis conjecture (ACC) \cite{AC} is 
a well-known open problem in combinatorial group theory deeply linked with the questions in low-dimensional topology.  
ACC states that every balanced presentation of the trivial group can be transformed into a trivial presentation by a sequence of simple transformations.


Several computational methods have been suggested for efficiently searching for such simplifications (e.g. \cite{Havas01andrews-curtisand, doi:10.1142/S0218196799000370,DBLP:journals/ijac/SwanOKE12,KS-2015, doi:10.1142/S0218196703001365, DBLP:conf/icms/Lisitsa18,shehper2024makesmathproblemshard}). However, there are still infinite sets of balanced trivial group presentations that could serve as potential counterexamples to the conjecture, for which the necessary simplifications have not yet been discovered.

For a group presentation  $\langle x_{1}, \ldots, x_{n}; r_{1}, \ldots r_{m} \rangle$ with generators $x_{i}$, 
and relators $r_{j}$, 
consider the following transformations. 

\begin{description}
\item[AC1] Replace some $r_{i}$ by $r_{i}^{-1}$. 
\item[AC2] Replace some $r_{i}$ by $r_{i}\cdot r_{j}$, $j\not=i$.
\item[AC3] Replace some $r_{i}$ by $w\cdot r_{i} \cdot w^{-1}$ where $w$ is any word in the generators. 
\item[AC4] Introduce a new generator $y$ and relator $y$ or delete a generator $y$ and relator $y$. 
\end{description} 

Two presentations $g$ and $g'$ are called \emph{Andrews-Curtis equivalent (AC-eq\-u\-i\-va\-lent)} if one of them can be obtained from the other by applying a finite sequence of transformations of the types (AC1) - (AC3). Two presentations are stably AC-equivalent if one of them can be obtained from the other by applying a finite
sequence of transformations of the types (AC1)–(AC4). A presentation $\langle x_{1}, \ldots, x_{n}; r_{1}, \ldots r_{m} \rangle$ is called \emph{balanced} if $n = m$. A balanced presentation is called \emph{AC-simplifiable}, or \emph{AC-trivializable}, if it is equivalent to trivial presentation $\langle x_{1}, \ldots, x_{n}; x_{1}, \ldots x_{n} \rangle$.

\begin{conjecture}[Andrews-Curtis \cite{AC}]

If $\langle x_{1}, \ldots, x_{n}; r_{1}, \ldots r_{n} \rangle$ is a balanced presentation of the trivial group it is AC-equivalent to the trivial presentation $\langle x_{1}, \ldots, x_{n}; x_{1}, \ldots x_{n} \rangle$\
\end{conjecture}


The stable version of the conjecture asserts that every balanced presentation of a trivial group is stably AC-equivalent (i.e., transformations using AC4 are permitted) to the trivial presentation. Both versions of the conjecture remain unresolved and present significant challenges.

\subsection{Akbulut-Kirby  presentations}

The infinite sequence of possible counterexamples was first introduced in \cite{key816520m}, where the authors defined an infinite family of balanced Akbulut-Kirby presentations of the trivial group: 
AK(n) = $\langle a,b \mid a^{n}b^{-(n+1)}, abab^{-1}a^{-1}b^{-1}\rangle$, $n\ge 2$.

The presentation AK(2) was shown to be AC-trivializable and, therefore, is not a counterexample, as demonstrated in \cite{doi:10.1142/S0218196799000370}. The AC-trivializability of all AK(n) for 
$n>2$ remains an open problem, with AK(3) being the shortest and perhaps most well-known potential counterexample to the AC conjecture. The same was true for the stable AC-trivializability of AK(n) for 
$n>2$ until August 2024, when it was shown in \cite{shehper2024makesmathproblemshard} that AK(3) is \emph{stably} trivializable. 
The result was established by identifying a sequence of "non-stable" AC-moves that transform a presentation, known from \cite{ACE} to be stably AC-simplifiable, into AK(3). This sequence was discovered using a greedy search algorithm, which was developed as part of a broader effort to study search and reinforcement learning algorithms in the context of the Andrews-Curtis conjecture, as detailed in \cite{shehper2024makesmathproblemshard}.



\subsection{Contributions}

In this communication, we present alternative proofs of the stable AC-trivializability of AK(3) using our approach based on automated deduction \cite{DBLP:journals/tinytocs/Lisitsa13,DBLP:conf/icms/Lisitsa18,
aitp-19} to identify the necessary reduction sequences. We provide five distinct transformation sequences, based on different subsets of AC-moves, each derived from two separate proofs obtained using the automated theorem prover Prover9 [12]. Additionally, we compare our solutions with that  in  \cite{shehper2024makesmathproblemshard} and highlight the challenge of developing computational methods for the efficient search of stable simplifications.

\section{Automated theorem proving for AC-simplifications}

In \cite{DBLP:journals/tinytocs/Lisitsa13,DBLP:conf/icms/Lisitsa18, aitp-19, LISITSA2024100168}, we developed an approach that utilizes automated deduction in first-order logic to search for trivializations, demonstrating that it is highly competitive. In this approach, we formalized AC-transformations—typically presented at the meta-level—at the object level, using term rewriting modulo group theory and first-order deduction. This section provides a brief overview of our approach.

Let $T_{G}$  be the equational theory of groups. 
For each  $n \ge 2$ we formulate~\cite{DBLP:conf/icms/Lisitsa18} a  natural term rewriting system modulo $T_{G}$, which captures AC-transformations of presentations of dimension $n$.  For $n=2$ that proceeds as follows.  
For an 
alphabet $A = \{a_{1}, a_{2}\}$
a  term rewriting system $ACT_{2}$ 
consists of the following rules: 

\begin{description}
\item[R1L] $f(x,y) \rightarrow f(r(x),y)$
\item[R1R] $f(x,y) \rightarrow f(x,r(y))$
\item[R2L] $f(x,y) \rightarrow f(x \cdot y,y)$  
\item[R2R] $f(x,y) \rightarrow f(x, y \cdot x)$
\item[R3L$_{i}$] $f(x,y) \rightarrow f((a_{i} \cdot x) \cdot r(a_{i}),y)$ for $a_{i} \in A, 
i = 1,2  $ 
\item[R3R$_{i}$] $f(x,y) \rightarrow f(x, (a_{i} \cdot y) \cdot r(a_{i}))$ for $a_{i} \in A, 
i = 1,2  $ 
\end{description}

The rewrite relations $\rightarrow_{ACT}$ for $ACT_{2}$ and  $\rightarrow_{ACT/G}$ for $ACT$ modulo theory $T_{G}$  on the set of all terms  are  defined 
 in the standard way \cite{Baader:1998:TR:280474}. The notion of  
rewrite relation 
$ \rightarrow_{ACT/G}$ 
captures adequately the 
notion of AC-rewriting, 
that is for presentations $p_{1}$ and $p_{2}$ we have 
$p_{1} \rightarrow_{AC}^{*} p_{2}$ iff $t_{p_{1}} 
\rightarrow_{ACT/G}^{*}.$ Here $t_{p}$ denotes a term encoding of a presentation $p$, that is for $p = \langle a_{1},a_{2} \mid t_{1}.t_{2}\rangle$ we have $t_{p} = f(t_{1},t_{2})$.

The term rewriting system $ACT_{2}$ can be 
 simplified  to the following reduced term rewriting system $rACT_{2}$~\cite{DBLP:conf/icms/Lisitsa18} without changing the transitive closure of the rewriting  relation: {\bf R1L}, {\bf R2L}, {\bf R2R}, {\bf R3L$_{1}$}, {\bf R3L$_{2}$}.



%

 In \cite{DBLP:conf/icms/Lisitsa18,LISITSA2024100168} two variants of translatiing   $ACT_{2}$ into first-order logic were explored: equational and implicational. In this communication, we focus solely on the implicational translation.
 
%


\noindent 

\subsection{Implicational Translation}~\label{imp-translation} 

Denote by $IG_{ACT_{2}}$ the first-order theory $T_{G} \cup rACT_{2}^{\rightarrow}$ where  $rACT_{2}^{\rightarrow}$ includes the following axioms: 
\begin{description}
\item[I-R1L] $R(x,y) \rightarrow R(r(x),y)$
\item[I-R2L] $R(x,y) \rightarrow R(x \cdot y,y)$  
\item[I-R2R] $R(x,y) \rightarrow R(x, y \cdot x)$
\item[I-R3L$_{i}$] $R(x,y) \rightarrow R((a_{i} \cdot x) \cdot r(a_{i}),y)$ for $a_{i} \in A, i = 1,2  $ 
\end{description}

\begin{proposition}~\label{prop:main} 
For ground terms $f(\bar{t}_{1})$ and $f(\bar{t}_{2})$, \\
$f(\bar{t}_{1})\rightarrow_{ACT_{2}/G}^{\ast} f(\bar{t}_{2})$ iff   $IG_{ACT_{2}} \vdash  R(\bar{t}_{1}) \rightarrow R(\bar{t}_{2})$
\end{proposition}

In a ``non-ground'' variant $IN_{ACT_{2}}$ of the implicational translation the axioms {\bf I-R3L$_{i}$} above  are replaced by a ``non-ground" axiom {\bf I-R3Z}: $R(x,y) \rightarrow R((z \cdot x) \cdot r(z),y)$  and the corresponding analogue of Proposition 1 holds true. 

Proposition~\ref{prop:main} (and its variants) underpins our automated deduction approach: in order to establish trivializability, or more generally AC-equivalence of groups presentations we delegate the task $IG_{ACT_{2}} \vdash^{?}  R(\bar{t}_{1}) \rightarrow R(\bar{t}_{2})$ to an automated theorem proving procedure (theorem prover). 

\begin{example} 
AK(4) is AC-simplifiable iff
$IG_{ACT_{2}} \vdash \varphi$  iff 
$IN_{ACT_{2}} \vdash \varphi$, where \\
$\varphi:=R((((((((a\cdot a) \cdot a) \cdot a) * r(b)) \cdot r(b)) \cdot r(b)) \cdot r(b)) * r(b),((((a \cdot b) \cdot a) \cdot r(b)) \cdot r(a)) \cdot r(b)) \rightarrow R(a,b)$.

\end{example}

%


\vspace*{-3mm} 
\section{Stable AC-trivializability of AK(3) by  automated deduction}

In their proof of stable AC-trivializability of AK(3) the authors of \cite{shehper2024makesmathproblemshard} have found a sequence of AC-transformations which transforms
balanced presentation {\scriptsize  P:=\\
$\langle a,b | a^{-1}b^{-1}ab^{-1}a^{-1}bab^{-2}aba^{-1}b, 
b^{-1}a^{-1}b^{2}a^{-1}b^{-1}abab^{-2}a\rangle$}
into AK(3). The presentation $P$ is a particular instance of a family of presentations for which it was shown in \cite{ACE} that they are all stably AC-trivializable.  By that stable AC-trivializability of AK(3) followed. 

While in \cite{shehper2024makesmathproblemshard} the transformation  sequence  was found by  an application of a {\em greedy search} algorithm we demontsratae here an alternative AC-transformation sequences found by automated theorem proving. 

For all proofs we have used automated theorem prover Prover9~\cite{prover9-mace4}.  We have published online all obtained proofs and extracted sequences\footnote{https://zenodo.org/records/14567743}. 

\subsection{First trivialization proof}

We applied Prover9 to the task of proving 
$IN_{ACT_{2}} \vdash R(\bar{t}_{1}) \rightarrow R(\bar{t}_{2})$, where $f(\bar{t}_{1}) = t_{P}$ is a term encoding of P, and $f(\bar{t}_{2}) = t_{AK(3)}$ is a term encoding  of AK(3).  In Prover9 syntax we have 

{\scriptsize
\begin{verbatim} 
Assumptions: 
<Group axioms for * (multiplication) and ' (inverse) > 
R(x,y) -> R(x',y).  
R(x,y) -> R(x,y * x). 
R(x,y) -> R(x * y,y). 
R(x,y) -> R((z * x) * z',y). 
Goal: 
R((((((((((((a' * b') * a) * b') * a') * b) * a) *
b') * b') * a) * b) * a') * b,((((((((((b' * a') * 
b) * b) * a') * b') * a) * b) * a) * b') * b') * a) -> 
R((((((a * a) * a) * b') * b') * b') * b',((((a * b) * 
a) * b') * a') * b')
\end{verbatim}
}
Note that $x'$ is used as shorthand  for $r(x)$, the inverse of $x$. 

The proof was completed  in 4273.70 seconds, establishing  the  AC-equivalence of P and AK(3).  The proof has 421 steps and the extracted sequence {\bf S1} of AC-transformations has length 252. 
Although the non-ground axiom {\bf I-R3Z}  was included in the assumptions, the actual proof only involves applications of this axiom, with 
z instantiated by the generators 
$a$, 
$b$, or their inverses $a'$ or $b'$. 
Every such application corresponds to an application of AC3 rule with a conjugation by a generator or an inverse generator.  We denote by {\bf S2}   a sequence of AC-tranformations  obtained from S1 by replacing all contiguous sequences of conjugation rules with their compositions, effectively simplifying them into single conjugations. {\bf  S2} has a length 160. We demontsrate it below, using the following notation convention. Inverted generators $a'$ and $b'$ are denoted by A and B respectively.  We use INV, MULT-L, MULT-R and CONJ wXw' to denote the applications of the  rules R1L, R2L, R2R and a version of R3L, that is conjugation by w, respectively.  

{\scriptsize
\begin{verbatim}
  1. ABaBAbaBBabAb,BAbbABabaBBa  -> INV 
  2. BaBAbbABabAba,BAbbABabaBBa  -> CONJ AbXBa
  3. BAbbABabAbaBa,BAbbABabaBBa  -> INV 
  4. AbABaBAbaBBab,BAbbABabaBBa  -> MULT-L
  5. AbABaaBBa,BAbbABabaBBa      -> INV 
  6. AbbAAbaBa,BAbbABabaBBa      -> CONJ BaXAb
  7. bAAba,BAbbABabaBBa          -> MULT-L 
  8. bAAbaBAbbABabaBBa,BAbbABabaBBa -> INV 
  9. AbbABAbaBBabABaaB,BAbbABabaBBa  -> MULT-R
 10. AbbABAbaBBabABaaB,ABaaB     -> INV 
 11. bAAbaBAbbABabaBBa,ABaaB     -> CONJ ABaaBXbAAba 
 12 .BAbbABabaBBabAAba,ABaaB     -> MULT-L 
 13. BAbbABabaBBa,ABaaB          -> CONJ BBabXBAbb
 14. ABabaBBaBAbb,ABaaB          -> INV 
 15. BBabAbbABAba,ABaaB          -> MULT-L 
 16. BBabAbbABaB,ABaaB           -> INV 
 17. bAbaBBaBAbb,ABaaB           -> CONJ ABaBxbAba
 18. BBaBAbbbAba,ABaaB           -> MULT-L 
 19. BBaBAbbbaB,ABaaB            -> CONJ bAbbXBBaB 
 20. AbbbaBBBaB,ABaaB            -> INV 
 21. bAbbbABBBa,ABaaB            -> MULT-L 
 22. bAbbbABBBBaaB,ABaaB         -> INV 
 23. bAAbbbbaBBBaB,ABaaB         -> CONJ aBXbA
 24. AbbbbaBBB,ABaaB             -> INV 
 25. bbbABBBBa,ABaaB             -> MULT-L 
 26. bbbABBBBBaaB,ABaaB          -> INV 
 27. bAAbbbbbaBBB,ABaaB          -> MULT-R
 28. bAAbbbbbaBBB,AbbbbaBBB      -> CONJ aBXbA
 29. AbbbbbaBBA,AbbbbaBBB        -> INV 
 30. abbABBBBBa,AbbbbaBBB        -> MULT-L 
 31. abbABaBBB,AbbbbaBBB         -> INV  
 32. bbbAbaBBA,AbbbbaBBB         -> CONJ AXa
 33. AbbbAbaBB,AbbbbaBBB         -> INV 
 34. bbABaBBBa,AbbbbaBBB         -> MULT-L 
 35. bbABabaBBB,AbbbbaBBB        -> CONJ baBBBXbbbAB
 36. baBABa,AbbbbaBBB            -> MULT-L 
 37. baBAbbbaBBB,AbbbbaBBB       -> INV 
 38. bbbABBBabAB,AbbbbaBBB       -> MULT-R 
 39. bbbABBBabAB,AbabAB          -> INV 
 40. baBAbbbaBBB,AbabAB          -> CONJ bABXbaB
 41. AbbbaBBaB,AbabAB            -> INV 
 42. bAbbABBBa,AbabAB            -> MULT-L 
 43. bAbbABBabAB,AbabAB          -> INV 
 44. baBAbbaBBaB,AbabAB          -> CONJ bABXbaB
 45. AbbaBBaaB,AbabAB            -> INV 
 46. bAAbbABBa,AbabAB            -> MULT-L 
 47. bAAbbABabAB,AbabAB          -> INV 
 48. baBAbaBBaaB,AbabAB          -> CONJ bABXbaB
 49. AbaBBaaaB,AbabAB            -> INV 
 50. bAAAbbABa,AbabAB            -> MULT-L 
 51. bAAAbbbAB,AbabAB            -> INV 
 52. baBBBaaaB,AbabAB            -> CONJ aaBXbAA
 53. aaaBBBa,AbabAB              -> MULT-L 
 54. aaaBBabAB,AbabAB            -> CONJ bABXbaB
 55. bABaaaBBa,AbabAB            -> MULT-L 
 56. bABaaaBabAB,AbabAB          -> CONJ bABXbaB
 57. bAABaaaBa,AbabAB            -> MULT-L 
 58. bAABaaaabAB,AbabAB          -> CONJ AbaaBXbAABa
 59. aaabAAABa,AbabAB            -> MULT-L 
 60. aaabAAbAB,AbabAB            -> CONJ AXa
 61. aabAAbABa,AbabAB            -> MULT-L 
 62. aabAAbbAB,AbabAB            -> CONJ AXa
 63. abAAbbABa,AbabAB            -> MULT-L 
 64. abAAbbbAB,AbabAB            -> CONJ AXa
 65. bAAbbbABa,AbabAB            -> MULT-L 
 66. bAAbbbbAB,AbabAB            -> CONJ aBXbA
 67. AbbbbAA,AbabAB              -> INV 
 68. aaBBBBa,AbabAB              -> MULT-L       
 69. aaBBBabAB,AbabAB            -> INV 
 70. baBAbbbAA,AbabAB            -> MULT-R 
 71. baBAbbbAA,AbbbbAA           -> CONJ bABXbaB
 72. AbbbAAbaB,AbbbbAA           -> INV 
 73. bABaaBBBa,AbbbbAA           -> MULT-L 
 74. bABaabAA,AbbbbAA            -> CONJ abAAXaaBA
 75. abAAbABa,AbbbbAA            -> MULT-L 
 76. abAAbAbbbAA,AbbbbAA         -> CONJ BaaBAXabAAb
 77. AbbbAbAAb,AbbbbAA           -> INV 
 78. BaaBaBBBa,AbbbbAA           -> MULT-L 
 79. BaaBabAA,AbbbbAA            -> CONJ AbXBa
 80. aBabAABa,AbbbbAA            -> MULT-L 
 81. aBabAAbbbAA,AbbbbAA         -> CONJ ABAbAXaBabA
 82. AbbbABabA,AbbbbAA           -> INV 
 83. aBAbaBBBa,AbbbbAA           -> MULT-L 
 84. aBAbabAA,AbbbbAA            -> CONJ abAAXaaBA
 85. abABAb,AbbbbAA              -> MULT-L 
 86. abABAbAbbbbAA,AbbbbAA       -> INV 
 87. aaBBBBaBabaBA,AbbbbAA       -> MULT-R 
 88. aaBBBBaBabaBA,BabaBA        -> INV 
 89. abABAbAbbbbAA,BabaBA        -> CONJ BabaBAXabABAb
 90. AbbbbAbABAb,BabaBA          -> MULT-L 
 91. AbbbbAA,BabaBA              -> CONJ BaXAb        
 92. bbbAAAb,BabaBA              -> MULT-L 
 93. bbbAAbaBA,BabaBA            -> CONJ aBAXabA
 94. aBAbbbAAb,BabaBA            -> MULT-L 
 95. aBAbbbAbaBA,BabaBA          -> CONJ aBAXabA
 96. aBBAbbbAb,BabaBA            -> MULT-L 
 97. aBBAbbbbaBA,BabaBA          -> CONJ BabbAXaBBAb  
 98. bbbaBBBAb,BabaBA            -> MULT-L 
 99. bbbaBBaBA,BabaBA            -> CONJ BXb
100. bbaBBaBAb,BabaBA            -> MULT-L 
101. bbaBBaaBA,BabaBA            -> CONJ BXb
102. baBBaaBAb,BabaBA            -> MULT-L 
103. baBBaaaBA,BabaBA            -> CONJ BXb
104. aBBaaaBAb,BabaBA            -> MULT-L 
105. aBBaaaaBA,BabaBA            -> CONJ bAXaB
106. BaaaaBB,BabaBA              -> INV 
107. bbAAAAb,BabaBA              -> MULT-L 
108. bbAAAbaBA,BabaBA            -> INV 
109. abABaaaBB,BabaBA            -> MULT-R 
110. abABaaaBB,BaaaaBB           -> CONJ aBAXabA
111. BaaaBBabA,BaaaaBB           -> INV 
112. aBAbbAAAb,BaaaaBB           -> MULT-L 
113. aBAbbaBB,BaaaaBB            -> CONJ baBBXbbAB
114. baBBaBAb,BaaaaBB            -> MULT-L 
115. baBBaBaaaBB,BaaaaBB         -> CONJ AbbABXbaBBa
116. BaaaBaBBa,BaaaaBB           -> INV 
117. AbbAbAAAb,BaaaaBB           -> MULT-L 
118. AbbAbaBB,BaaaaBB            -> CONJ BaXAb
119. bAbaBBAb,BaaaaBB            -> MULT-L 
120. bAbaBBaaaBB,BaaaaBB         -> CONJ bABaBXbAbaB
121. BaaaBAbaB,BaaaaBB           -> INV 
122. bABabAAAb,BaaaaBB           -> MULT-L 
123. bABabaBB,BaaaaBB            -> CONJ baBXbAB
124. abaBAB,BaaaaBB              -> INV 
125. babABA,BaaaaBB              -> MULT-L 
126. babABABaaaaBB,BaaaaBB       -> INV 
127. bbAAAAbabaBAB,BaaaaBB       -> MULT-R
128. bbAAAAbabaBAB,abaBAB        -> INV 
129. babABABaaaaBB,abaBAB        -> CONJ abaBABXbabABA
130. BaaaaBabABA,abaBAB          -> MULT-L 
131. BaaaaBB,abaBAB              -> CONJ bXB
132. aaaaBBB,abaBAB              -> INV 
133. bbbAAAA,abaBAB              -> MULT-L 
134. bbbAAAbaBAB,abaBAB          -> INV 
135. babABaaaBBB,abaBAB          -> CONJ BXb
136. abABaaaBB,abaBAB            -> INV 
137. bbAAAbaBA,abaBAB            -> MULT-L 
138. bbAAAbaaBAB,abaBAB          -> INV 
139. babAABaaaBB,abaBAB          -> CONJ BXb
140. abAABaaaB,abaBAB            -> INV 
141. bAAAbaaBA,abaBAB            -> MULT-L 
142. bAAAbaaaBAB,abaBAB          -> INV 
143. babAAABaaaB,abaBAB          -> CONJ BXb
144. abAAABaaa,abaBAB            -> INV 
145. AAAbaaaBA,abaBAB            -> MULT-L 
146. AAAbaaaaBAB,abaBAB          -> INV 
147. babAAAABaaa,abaBAB          -> CONJ aXA
148. ababAAAABaa,abaBAB          -> INV 
149. AAbaaaaBABA,abaBAB          -> MULT-L 
150. AAbaaaaBBAB,abaBAB          -> INV 
151. babbAAAABaa,abaBAB          -> CONJ aXA
152. ababbAAAABa,abaBAB          -> INV 
153. AbaaaaBBABA,abaBAB          -> MULT-L 
154. AbaaaaBBBAB,abaBAB          -> INV 
155. babbbAAAABa,abaBAB          -> CONJ aXA
156. ababbbAAAAB,abaBAB          -> INV 
157. baaaaBBBABA,abaBAB          -> MULT-L 
158. baaaaBBBBAB,abaBAB          -> INV 
159. babbbbAAAAB,abaBAB          -> CONJ ABXba
160. bbbbAAA,abaBAB              -> INV 
     aaaBBBB,abaBAB
\end{verbatim} 
}

A visual inspection of all the presentations included in {\bf S2} 
  suggests at least two possible ways to reduce the length of the transformation sequence by incorporating additional types of elementary transformations beyond those allowed by  $rAST_{2}$.

We denote by {\bf S3} a sequence of transformations which mathches S2 from positions 1 to 67 and  is followed by 
{\scriptsize
\begin{verbatim}
68. aaBBBBa,AbabAB              -> CONJ aXA
69. aaaBBBB,AbabAB              -> CONJ-R aXA
70  aaaBBBB,babABA              -> INV-R
    aaaBBBB,abaBAB
\end{verbatim} 
}

Here CONJ-R and INV-R  denote a conjugation  and inversion applied to the second (right) component of a presentation, respectively. 

Similarly by {\bf S4}  we denote a sequence which matches  {\bf S2}  from positions 1 to 52  and is followed by 
{\scriptsize
\begin{verbatim} 
 53. aaaBBBa,AbabAB              -> Aut[a->b,b->a)] 
 54. bbbAAAb,BabaBA              -> CONJ bXB
 55. bbbbAAA,BabaBA              -> INV 
 56. aaaBBBB,BabaBA              -> CONJ-R bXB
 57. aaaBBBB,abaBAB
\end{verbatim} 
}

The notation \verb"Aut[a->b,b->a]"
 represents an automorphism of the free group 
that swaps the two generators. It is straightforward to demonstrate that applying automorphism transformations does not alter the AC-equivalence class when applied to presentations that are AC-equivalent to the trivial one. In \cite{PU16}, it was also shown that this holds for presentations AC-equivalent to any of 
AK(n), $n > 2$.

In summary, from the first proof, we have derived four sequences of AC moves that transform the presentation P into 
AK(3). The lengths of these sequences are 252, 160, 70, and 57, respectively, each using different subsets of elementary transformations.

\subsection{Second proof} 

The original sequence of moves transforming P to AK(3) found in \cite{shehper2024makesmathproblemshard} has the length 53 which is less than the length of any of the sequences we have extracted from the first proof. 
However, it is worth noting that the set of elementary AC-moves discussed in \cite{shehper2024makesmathproblemshard} differs from the standard moves we have covered so far. This set consists of the instantiations of the following templates:

\begin{description}
\item[AC'1] Replace some $r_{i}$ by $r_{i}r_{j}^{\pm 1}$ for for $i\not=j$
\item[AC'2] Change some $r_{i}$ to $gr_{i}g^{-1}$ where $g$ is a generator or its inverse.
\end{description}

In the case of balanced presentations with two generators, the templates {\bf [AC'1]} and {\bf [AC'2]} give rise to 12 distinct rules. We can easily adapt the implicational translation discussed in subsection \ref{imp-translation} to this case, ensuring that the analogue of Proposition 1 holds true. The logic translations of these 12 rules in Prover9 syntax are:

{\scriptsize
\begin{verbatim}
R(x,y) -> R(x,y * x).           R(x,y) -> R(x,(b' * y) * b).
R(x,y) -> R(x * y',y).          R(x,y) -> R((a * x) * a',y). 
R(x,y) -> R(x,y * x').          R(x,y) -> R(x,(a * y) * a'). 
R(x,y) -> R(x * y,y).           R(x,y) -> R((b * x) * b',y).  
R(x,y) -> R(x,(a' * y) * a).    R(x,y) -> R(x,(b * y) * b').  
R(x,y) -> R((b' * x) * b,y).    R(x,y) -> R((a' * x) * a,y).  
 \end{verbatim} 
}

We denote by IMG the first-order theory $T_{G} \cup MR$ where $MR$ consists of the above 12 axioms encoding the modified rules. We have applied Prover9 to the task of proving  the same goal  as  in the first proof  from IMG. 
The proof was completed in 927.61 seconds establishing yet again AC-equivalence of presentations P and AK(3). 
The proof has a length of 158, while the length of the extracted sequence {\bf S5}  of AC-transformations (applications of one of the 12 rules) is 110. Its length is more than double that of the original sequence of the same rules produced in \cite{shehper2024makesmathproblemshard}.
Thus, we were unable to answer the open question in \cite{shehper2024makesmathproblemshard} about whether the sequence of length 53 found there is the shortest possible sequence transforming
P into  AK(3).

\subsection{Implementation details} 

For all experiments we have used LADR-Dec-2007 version of Prover9~\cite{prover9-mace4} running on Intel(R) Core(TM) i5-7200U CPU @ 2.50GHz   2.71 GHz, 16 GB RAM, Windows 10 Enterprise. 


\section{Conclusion} 

We have presented alternative proofs of the stable AC-trivializability of 
AK(3), based on the use of automated deduction. We  presented five  sequences of AC-transformations derived from two different proofs demonstrating the flexibility of our method which can be adapted to encode different variants of rules. 
All of these may be of interest in the study of AC-reachability in general.
These results also support our claim, made in \cite{DBLP:journals/tinytocs/Lisitsa13,DBLP:conf/icms/Lisitsa18}, that for all proposed potential counterexamples where non-trivial AC-simplifications were found using specialized search algorithms, AC-simplifications can also be discovered using generic theorem proving techniques\footnote{On the other hand, novel AC-simplifications achieved through automated deduction can be found in \cite{DBLP:conf/icms/Lisitsa18, LISITSA2024100168}}.

Finally, we note that in both \cite{shehper2024makesmathproblemshard} and this communication, the search conducted to establish the result on stable AC-simplifiability was actually focused on non-stable transformations. We are not aware of any computational approach that can directly search for stable AC-transformations. Our experiments with natural encodings of this task in logic, along with the use of automated theorem provers, have yet to yield any significant results. We conclude by proposing a challenge to develop computational methods for searching stable AC-transformations.

\section{Declaration of generative AI and AI-assisted technologies in the writing process} 

During the preparation of this work the author used ChatGPT 4o mini service  in order to improve readability of the manuscript. After using this service, the author reviewed and edited the content as needed and takes full responsibility for the content of the published article.

\bibliographystyle{plain}
\bibliography{AC.bib}

\begin{thebibliography}{10}

\bibitem{key816520m}
Selman Akbulut and Robion Kirby.
\newblock A potential smooth counterexample in dimension {4} to the {P}oincar\'e conjecture, the {S}choenflies conjecture, and the {A}ndrews--{C}urtis conjecture.
\newblock {\em Topology}, 24(4):375--390, 1985.
\newblock Available at http://dx.doi.org/10.1016/0040-9383(85)90010-2. MR 87d:57024. Zbl 0584.57009.

\bibitem{ACE}
Alexei G.~Myasnikov Alexei D.~Myasnikov and Vladimir Shpilrain.
\newblock On the {Andrews–Curtis} equivalence.
\newblock {\em Contemporary Mathematics}, 296, 2002.

\bibitem{AC}
J.~Andrews and M.L. Curtis.
\newblock Free groups and handlebodies.
\newblock {\em Proc. Amer. Math. Soc.}, 16:192--195, 1965.

\bibitem{Baader:1998:TR:280474}
Franz Baader and Tobias Nipkow.
\newblock {\em Term Rewriting and All That}.
\newblock Cambridge University Press, New York, NY, USA, 1998.

\bibitem{Havas01andrews-curtisand}
George Havas and Colin Ramsay.
\newblock {Andrews-Curtis} and {Todd-Coxeter} proof words.
\newblock Technical report, in Oxford. Vol. I, London Math. Soc. Lecture Note Ser, 2001.

\bibitem{doi:10.1142/S0218196703001365}
George Havas and Colin Ramsay.
\newblock Breadth-first search and the {Andrews-Curtis} conjecture.
\newblock {\em International Journal of Algebra and Computation}, 13(01):61--68, 2003.

\bibitem{KS-2015}
Krzysztof Krawiec and Jerry Swan.
\newblock {AC-trivialization proofs eliminating some potential counterexamples to the Andrews-Curtis conjecture}.
\newblock Available at www.cs.put.poznan.pl/kkrawiec/wiki/uploads/ Site/ACsequences.pdf, 2015.

\bibitem{DBLP:journals/tinytocs/Lisitsa13}
Alexei Lisitsa.
\newblock First-order theorem proving in the exploration of {Andrews-Curtis} conjecture.
\newblock {\em TinyToCS}, 2, 2013.

\bibitem{DBLP:conf/icms/Lisitsa18}
Alexei Lisitsa.
\newblock {The Andrews-Curtis Conjecture, Term Rewriting and First-Order Proofs}.
\newblock In {\em Mathematical Software - {ICMS} 2018 - 6th International Conference, South Bend, IN, USA, July 24-27, 2018, Proceedings}, pages 343--351, 2018.

\bibitem{aitp-19}
Alexei Lisitsa.
\newblock Automated reasoning for the {Andrews-Curtis} conjecture.
\newblock In {\em AITP 2019, Fourth Conference on Artificial Intelligence and Theorem Proving, Abstracts of the Talks April 7–12, 2019, Obergurgl, Austria}, pages 82--83, 2019.

\bibitem{LISITSA2024100168}
Alexei Lisitsa.
\newblock New {Andrews–Curtis} trivializations for {Miller–Schupp} group presentations.
\newblock {\em Examples and Counterexamples}, 6:100168, 2024.

\bibitem{prover9-mace4}
W.~McCune.
\newblock Prover9 and mace4.
\newblock \verb|http://www.cs.unm.edu/~mccune/prover9/|, 2005--2010.

\bibitem{doi:10.1142/S0218196799000370}
Alexei~D. Miasnikov.
\newblock Genetic algorithms and the {Andrews-Curtis} conjecture.
\newblock {\em International Journal of Algebra and Computation}, 09(06):671--686, 1999.

\bibitem{PU16}
Dmitry Panteleev and Alexander Ushakov.
\newblock Conjugacy search problem and the {Andrews-Curtis conjecture}.
\newblock arxiv:1609.00325, 2016.

\bibitem{shehper2024makesmathproblemshard}
Ali Shehper, Anibal~M. Medina-Mardones, Bartłomiej Lewandowski, Angus Gruen, Piotr Kucharski, and Sergei Gukov.
\newblock What makes math problems hard for reinforcement learning: a case study.
\newblock arxiv:2408.15332, 2024.

\bibitem{DBLP:journals/ijac/SwanOKE12}
Jerry Swan, Gabriela Ochoa, Graham Kendall, and Martin Edjvet.
\newblock {Fitness Landscapes and the Andrews-Curtis Conjecture}.
\newblock {\em {IJAC}}, 22(2), 2012.

\end{thebibliography}

\end{document}